\documentclass[journal,twoside]{IEEEtranTCOM}

\usepackage{blindtext}

\usepackage[latin1]{inputenc}

\usepackage[noadjust]{cite}
\usepackage{graphicx}
\usepackage{psfrag}
\usepackage{subfig}
\usepackage{amsmath,amssymb,amsthm,mathrsfs,amsfonts,dsfont}
\usepackage{color}
\usepackage{colortbl}
\usepackage[algoruled]{algorithm2e}

\usepackage{fixltx2e}

\usepackage{multirow}
\usepackage{multicol}
\usepackage[hidelinks]{hyperref}

\usepackage{amsthm}
\newtheoremstyle{note}{3pt}{3pt}{\itshape}{}{\itshape}{:}{.5em}{}
\theoremstyle{note}

\usepackage{acronym}
\newacro{PLC}{power line communication}
\newacro{OFDM}{orthogonal frequency-division multiplexing}
\newacro{HS-OFDM}{Hermitian symmetric OFDM}
\newacro{PSD}{power spectral density}
\newacro{BER}{bit error rate}
\newacro{SER}{symbol error rate}
\newacroindefinite{SER}{an}{a}
\newacro{SCRA}{spectral compressive resource allocation}
\newacro{TCRA}{temporal compressive resource allocation}
\newacro{STCRA}{spectral-temporal compressive resource allocation}
\newacro{nSNR}{normalized signal-to-noise ratio}
\newacroindefinite{nSNR}{an}{a}
\newacro{CFR}{channel frequency response}
\newacro{LTV}{linear time-variant}
\newacro{LPTV}{linear periodically time variant}
\newacroindefinite{LPTV}{an}{a}
\newacro{LTI}{linear time-invariant}
\newacroindefinite{LTI}{an}{a}
\newacro{AWGN}{additive white Gaussian noise}
\newacro{QAM}{quadrature amplitude modulation}
\newacro{SNR}{signal-to-noise ratio}
\newacroindefinite{SNR}{an}{a}
\newacro{FT}{Fourier transform}
\newacro{CCDF}{complementary cumulative distribution function}
\newacro{CIR}{channel impulse response}
\newacro{AIC}{Akaike information criterion}
\newacro{BIC}{Bayesian information criterion}
\newacro{EDC}{efficient determination criterion}
\newacro{PDF}{probability density function}
\newacro{GMD}{Gaussian mixture distribution \acused{GMDs}}
\newacro{GMDs}{Gaussian mixture distributions \acused{GMD}}
\newacro{GC}{Gaussian component}
\newacro{CP}{component proportion}
\newacro{CSI}{channel state information}

\newacro{FDM}{frequency-division multiplexing}
\newacro{NBI}{narrowband interference}
\newacro{DSL}{digital subscriber line}
\newacro{RA}{rate-adaptive}
\newacro{MA}{margin-adaptive}
\newacro{DMT}{discrete multitone modulation}
\newacro{ZF}{zero-forcing}
\newacro{PAM}{pulse amplitude modulation}
\newacro{WF}{water filling}
\newacro{WSS}{wide-sense stationary}
\newacro{ISI}{intersymbol interference}
\newacro{QoS}{quality of service}

\newacro{EMC}{electromagnetic compatibility}

\newacro{TL}{transmission line}
\newacro{DL}{distribution line}
\newacro{HV}{high voltage}
\newacro{MV}{medium voltage}
\newacro{LV}{low voltage}
\newacro{TEM}{transverse electromagnetic}
\newacro{SG}{smart grid}
\newacro{DFT}{discrete Fourier transform}
\newacro{IDFT}{inverse discrete Fourier transform}
\newacro{FFT}{fast Fourier transform}
\newacro{IFFT}{inverse fast Fourier transform}
\newacro{IFT}{inverse Fourier transform}
\newacro{HIF}{high impedance fault}
\newacro{LIF}{low impedance fault}
\newacro{TDR}{time-domain reflectometry}
\newacro{FDR}{frequency-domain reflectometry}
\newacro{JTFDR}{joint time-frequency domain reflectometry}
\newacro{MCR}{multicarrier reflectometry}
\newacro{MCTDR}{multicarrier time-domain reflectometry}
\newacro{OMTDR}{orthogonal multitone time-domain reflectometry}

\newacro{IDTFT}{inverse discrete-time Fourier transform}
\newacro{PLM}{power line modem}
\newacro{DFnT}{discrete Fresnel transform}
\newacro{OCDM}{orthogonal chirp-division multiplexing}
\newacro{PSLR}{peak-to-sidelobe level ratio}
\newacro{ISLR}{integrated-sidelobe level ratio}

\newacro{MIMO}{multiple-input multiple-output}
\newacro{TDMA}{time-division multiple access}
\newacro{FDMA}{frequency-division multiple access}
\newacro{CDMA}{code-division multiple access}
\newacro{HS-OFDMA}{Hermitian symmetric orthogonal frequency-division multiple access}

\newacro{SINR}{signal-to-interference-plus-noise ratio}
\newacro{NB}{narrowband}
\newacro{BB}{broadband}

\newacro{BPSK}{binary phase-shift keying}
\newacro{QPSK}{quadrature phase-shift keying}
\newacro{8PSK}{eight phase-shift keying}

\newacro{FCC}{Federal Communications Commission}
\newacro{ARIB}{Association of Radio Industries and Businesses}
\newacro{CENELEC}{European Committee for Electrotechnical Standardization (\textit{Comit\'e Europ\'en de Normalisation \'Electrotechnique} in French)}

\newacro{MTL}{multiconductor transmission line}

\usepackage{dblfloatfix}
\newcommand{\Rho}{\mathrm{P}}

\usepackage{accents}

\usepackage{multirow}
\usepackage[table,xcdraw]{xcolor}

\usepackage{flushend}

\begin{document}
\title{HS-OFDM-based Time-Domain Reflectometry for Power Line Sensing: Characteristics and Limitations}

\author{Lucas Giroto de Oliveira,~\IEEEmembership{Student Member,~IEEE},
	Mateus de L. Filomeno,\\ and Mois\'es V. Ribeiro,~\IEEEmembership{Senior Member,~IEEE}
	\thanks{Manuscript received MM DD, YYYY; revised MM DD, YYYY. This study was financed in part by the Coordena\c{c}\~ao de Aperfei\c{c}oamento de Pessoal de N\'ivel Superior - Brasil (CAPES) - Finance Code 001, Copel Distribui\c{c}\~ao LTD - PD 2866-0420/2015 , CNPq, FAPEMIG, INERGE and Smarti9 LTD.}
	\thanks{Lucas Giroto de Oliveira and Mateus de L. Filomeno are with the Electrical Engineering Department, Federal University of Juiz de Fora (UFJF), Juiz de Fora, Brazil  (e-mail: lgiroto@ieee.org, mateus.lima@engenharia.ufjf.br).}
	\thanks{Mois\'es V. Ribeiro is with the Electrical Engineering Department, Federal University of Juiz de Fora (UFJF), Juiz de Fora, Brazil, and Smarti9 Ltd, Brazil  (e-mail: mribeiro@ieee.org).}}


\maketitle

\begin{abstract}
	This study discusses key characteristics and limitations of time-domain reflectometry (TDR) systems based on the Hermitian symmetric orthogonal frequency-division multiplexing (HS-OFDM) scheme for power line sensing. In this sense, a system model with a power line modem injecting signals and capturing raising reflections for sensing a power distribution network is outlined. Next, pulse compression and channel estimation reflectogram processing approaches are carefully described and the effects of system parametrization and multiple access schemes on the HS-OFDM-based TDR system performance are addressed. Finally, numerical results covering a comparison between pulse compression and channel estimation, system limitations based on parametrization considering typical European underground low-voltage and US overhead medium-voltage (MV) scenarios and narrowband (NB) power line communication (PLC) regulatory constraints, and comparison among multiple access techniques in a Brazilian MV overhead scenario are presented for supporting the carried out discussion. Based on the attained results, it is shown that channel estimation outperforms the pulse compression in terms of computational complexity and sidelobe level. Also, it is shown that the NB-PLC frequency range provides fair range resolution and maximum unambiguous range values. Finally, it is seen that the use of the frequency-division multiple access multiple access schemes presents different signal-to-interference-plus-noise ratio (SINR) performance among different power line modems (PLMs) connected to a power distribution grid, while the use of time-division multiple access and code-division multiple access schemes results in fair SINR performance among the PLMs at the cost of obtaining  less reflectograms over time due to time multiplexing and spreading processes, respectively.
\end{abstract}


\IEEEpeerreviewmaketitle

\section{Introduction}\label{sec:introduction}


Today, electrical energy is a key resource in the entire world. The need to connect power generating units to final consumers has led to a significant expansion of power transmission and distribution infrastructures. Such expansion implies in greater subjectivity of the latter to faults along with both infrastructures. In order to minimize the interruption time as well as the damages resulting from faults, several protection schemes have been developed.

Conventional protection schemes for power line networks, however, fail in tasks such as detection and location of hard faults, e.g. \acp{HIF} or distant \acp{LIF}, and soft faults, eg. cable degradation. Among more efficient alternatives are analysis harmonic currents and voltages \cite{sedighizadeh2010,ghaderi2017}, and the use of traveling waves for detecting and locating impedance discontinuities along the network. The latter, which has long been subject of study for sensing of wired networks in scenarios such as aircraft \cite{auzanneau2013}, has been gaining attention for sensing of power lines. In this context, the adaptation of \acp{PLM} for network sensing \cite{passerini2018_2} as well as techniques that exploit characteristics of traveling waves propagating back and forward the power line for applications such as topology inference \cite{ahmed2013} or fault detection and location \cite{milioudis2015} have been investigated.

Potential techniques for wired network sensing based on traveling waves are impedance spectroscopy, transferometry, and reflectometry \cite{auzanneau2013,auzanneau2016}. Variations of the latter are \ac{TDR}, \ac{FDR} \cite{furse2006}, and \ac{JTFDR} \cite{wang2010}, which consist of injecting signals into the analyzed network and capturing the raised reflections, which are respectively processed in time, frequency, or both domains. The use of \ac{TDR} principles for fault sensing in power lines, subject of study in the past \cite{taylor1996,bo1999}, has been recently revisited in the literature \cite{paulis2016} due to low computational complexity associated with the post-processing of obtained reflectograms. Such aspect of \ac{TDR}-based techniques allows on-line obtaining of reflectograms \cite{hassen2015}, therefore enabling efficient sensing of power line networks at both symbol and mains levels \cite{passerini2018_2}.

Effectiveness in terms of digital signal processing and \ac{EMC} without the inherent drawback of poor spectrum control in classical \ac{TDR} techniques can be achieved by \ac{MCTDR} \cite{naik2006,amini2009} and its time-domain version \cite{lelong2009}, which only differ in the domain where reflectogram processing takes place. For enhanced compatibility with existing \ac{PLC} systems an optimized digital signal processing, an interesting approach is the \ac{OMTDR} \cite{hassen2015}, which is variation of the \ac{MCTDR} based on \ac{OFDM}.

Although \ac{TDR} based on \ac{OFDM} or its baseband version \ac{HS-OFDM} seems a good candidate for power line sensing, aspects such as measurement range and number of reflectograms obtained over time become more relevant due to the long distances in power distribution networks and the \ac{LPTV} behavior of transfer functions experienced by signals injected by \acp{PLM} into the network. To the best of the author's knowledge, the use of channel estimation, performed for communication purposes in \ac{PLC} systems \cite{oliveira2014}, rather than the usual pulse compression performed for reflectometric sensing in wired networks \cite{chang2015,hassen2015} has also not been addressed in the literature. For an efficient distributed sensing, the effect of multiple access on \ac{TDR} measurements, already considered for pulse compression \ac{TDR} in some aspects \cite{lelong2010,hassen2015}, must also be taken into account for \ac{HS-OFDM}-based \ac{TDR} systems relying on channel estimation. Given this context, the present study investigates the influences of the parametrization of an \ac{HS-OFDM}-based \ac{TDR} system as well as limitations imposed by multiple access schemes on its performance for power line sensing.

The main contributions of this study are summarized as follows.
\begin{enumerate}
	\item Description of an \ac{HS-OFDM}-based \ac{TDR} system sensing a power distribution network. Based on this formulation, we discuss the pulse compression and channel estimation reflectogram processing approaches and analyze the effect of the \ac{TDR} system parametrization on its performance using \ac{PSLR}, \ac{ISLR}, range resolution, and maximum unambiguous range as metrics taking into account regulatory constraints and considering typical European underground \ac{LV} and US overhead \ac{MV} scenarios.
	\item Introduction of \ac{TDMA}, \ac{FDMA}, and \ac{CDMA} multiple access schemes for an \ac{HS-OFDM}-based \ac{TDR} system for power line sensing based on the channel estimation procedure and comparative analysis among them in terms of number of obtained reflectograms and \ac{SINR} considering a Brazilian overhead \ac{MV} scenario.
\end{enumerate}
Our major findings are as follows:
\begin{enumerate}
	\item Channel estimation outperforms the pulse compression procedure in terms of both computational complexity and sidelobe level, being therefore a more attractive reflectogram processing approach for \ac{HS-OFDM}-based \ac{TDR} systems.
	\item \ac{HS-OFDM}-based \ac{TDR} systems operating in the \ac{NB}-\ac{PLC} frequency range are suitable for the sensing of \ac{LV} and \ac{MV} power distribution networks sections, offering fair range resolution and maximum unambiguous range values.
	\item The use of \ac{TDMA} and \ac{CDMA} multiple access schemes results in a smaller number of obtained reflectograms over time respectively due to time multiplexing and spreading processes and in a fair \ac{SINR} level among the multiple \acp{PLM} connected to the power distribution network, being the \ac{SINR} higher in the \ac{CDMA} case due to \ac{SNR} gain provided by this scheme. On the other hand, the use of \ac{FDMA} results in a higher number of obtained reflectograms over time in comparison with \ac{TDMA} and \ac{CDMA} and in higher \ac{SINR} to \acp{PLM} associated with subcarriers in higher frequnecy bins due to the exponentially decreasing additive noise \ac{PSD}.
\end{enumerate}

The remainder of this paper is organized as follows. Section~\ref{sec:model} describes the \ac{HS-OFDM}-based \ac{TDR} system for sensing of a power distribution network. Section~\ref{sec:processing} discusses the pulse compression and channel estimation procedures for reflectogram processing. Next, Section~\ref{sec:limitations} addresses the effect of system parametrization on its performance and Section~\ref{sec:multiple_access} discusses the use of multiple access schemes for power line sensing with multiple \acp{PLM}. A numerical analysis for supporting the carried out discussion is carried out in Section~\ref{sec:analysis}. Finally, concluding remarks are placed in Section~\ref{sec:conclusion}.

\subsection*{Notation}\label{subsec:notation}

Throughout the paper, $(\cdot)^T$ and $(\cdot)^\dagger$ indicate the transpose and Hermitian transpose operators, respectively; $\star$ is the convolution operator; $\odot$ and $\oslash$ denote the Hadamard product and division, which respectively perform element-wise multiplication and division of two equal-sized matrices; $\mathbb{E}\{\cdot\}$ represents the expectation operator; the $M$-size \ac{DFT} matrix is denoted by $\mathbf{W}_{M}$; $\mathbf{0}_{a\times b}$ and $\mathbf{1}_{a\times b}$ denote $(a\times b)$-size matrices respectively composed of zeros and ones.

\section{System Model}\label{sec:model}

%
%
Let a baseband \ac{TDR} system be consisted by a full-duplex \ac{PLM} connected to a power distribution network, which, at a single given point, injects signals and captures reflections that travel at a phase velocity $v_p$. Assuming that the injection and subsequent capture of reflections of signals takes place within a coherence time $T_c$, in which variations in loads or any other element of the network are irrelevant, one can consider the power distribution network as \iac{LTI} system. The reflections captured by the \ac{PLM} from such power distribution network, which are raised by impedance discontinuities along the path traveled by the injected signal, are therefore the output of a reflection channel with impulse response $h_\Gamma(t)$.

Such reflections are quantified via the input reflection coefficient between the \ac{PLM} of output impedance denoted in the continuous frequency-domain as $Z_{PLM}(f)$ and the power distribution network of input impedance $Z_{in}(f)$, being expressed as \cite{paul2007}
\begin{equation}\label{eq:Gamma_in}
\Gamma_{in}(f) = \frac{Z_{in}(f)-Z_{PLM}(f)}{Z_{in}(f)+Z_{PLM}(f)}.
\end{equation}
From $\Gamma_{in}(f)$, one finally obtains the impulse response of the reflection channel via the inverse Fourier Transform, i.e.,
\begin{equation}\label{eq:h(t)}
h_\Gamma(t) = \int_{-\infty}^{\infty}\Gamma_{in}(f)e^{j2\pi ft}df.
\end{equation}

The reflectometric sensing of the power distribution network can therefore be performed via an analysis of the reflections raised by the impedance discontinuities. For this purpose, the considered \ac{TDR} system is used for obtaining a reflectogram, which is an estimate of the reflection channel impulse response. In this paper, it is assumed that such procedure is performed by an \ac{HS-OFDM} system that is band-limited to a bandwidth $B$ and has sampling frequency $F_s=2B$.

\begin{figure*}[!t]
	\centering	
	\psfrag{a}[c][c]{$\mathbf{D}$}
	\psfrag{b}[c][c]{$\mathbf{s}$}
	\psfrag{c}[c][c]{$\mathbf{r}$}
	\psfrag{d}[c][c]{$\boldsymbol{\rho}$}
	\psfrag{P}[c][c]{$\mathcal{P}(\cdot)$}
	\psfrag{Q}[c][c]{$\mathcal{Q}(\cdot)$}
	\psfrag{s}[c][c]{$s(t)$}
	\psfrag{h}[c][c]{$h_\Gamma(t)$}
	\psfrag{r}[c][c]{$\tilde{r}(t)$}
	\psfrag{n}[c][c]{$v(t)$}
	\psfrag{y}[c][c]{$r(t)$}
	\includegraphics[height=5.5cm]{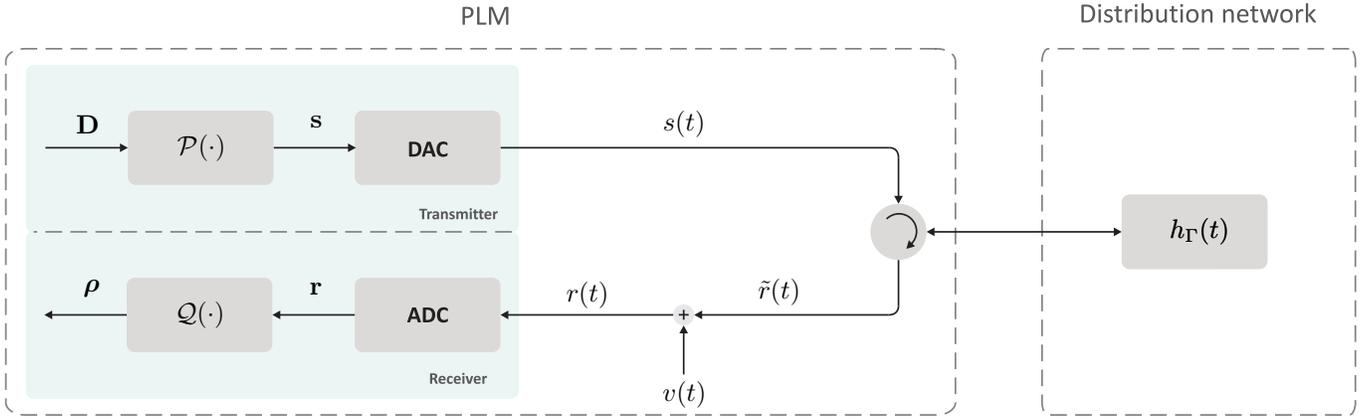}
	\caption{HS-OFDM-based TDR system over a power distribution network.}\label{fig:OFDM_TDR}	
\end{figure*}

The considered \ac{HS-OFDM}-based \ac{TDR} system is depicted in Fig.~\ref{fig:OFDM_TDR}, starting with a complex vector \mbox{$\mathbf{D}=[D_0,D_1,\cdots,D_{N-1}]^T$}, such that $\mathbf{D}\in\mathbb{C}^{N\times1}$. This vector is inputted to the function $\mathcal{P}(\cdot)$ that represents, in a condensed form, the digital processing performed at the transmitter side, which is composed by three processing stages. The first one is the Hermitian symmetric mapping $Map(\cdot)$ \cite{giroto2018} that transforms $\mathbf{D}$ into the $2N$-length discrete-frequency domain \ac{HS-OFDM} symbol \mbox{$\mathbf{X}=[X_0,X_1,\cdots,X_{2N-1}]^T$} such that $\mathbf{X}\in\mathbb{C}^{2N\times1}$. Next, the \ac{IDFT} transforms $\mathbf{X}$ into the discrete-time domain vector $\mathbf{x}\in\mathbb{R}^{2N\times1}$, which is real-valued due to the Hermitian symmetry of $\mathbf{X}$. This is expressed as $\mathbf{x}=\frac{1}{\sqrt{2N}}\mathbf{W}_{2N}^\dagger\mathbf{X}$. Finally, an $L_{cp}$-length cyclic prefix is appended to $\mathbf{x}$, resulting in the vector $\mathbf{s}\in\mathbb{R}^{(2N+L_{cp})\times1}$. Considering that the reflection channel impulse response has an $L_h$-length discrete-time domain representation $\mathbf{h}_\Gamma\in\mathbb{R}^{L_h\times1}$, no \ac{ISI} is experienced if the constraint $L_{cp}\leq L_h$ is satisfied.

Back to Fig.~\ref{fig:OFDM_TDR}, the discrete-time vector $\mathbf{s}$ undergoes an digital-to-analog conversion, being converted into the continuous-time signal $s(t)$ that is inputted to the reflection channel of impulse response $h_\Gamma(t)$. It is worth highlighting that, in order for the \ac{LTI} assumption of the reflection channel to hold, the duration of $s(t)$, i.e., the \ac{HS-OFDM} symbol in the continuous-time domain, $T_{symb}=(2N+L_{cp})T_s$ must satisfy the constraint $T_{symb}\ll T_c$. The first term of the previous expression accounts for the number of samples of the discrete-time vector $\mathbf{s}$, while $T_s=1/F_s$ is the sampling period. The resulting signal from the convolution between the transmit signal and the channel is \mbox{$\tilde{r}(t)=s(t)\star h_\Gamma(t)$}. To this signal is added the noise $v(t)$, which is a zero-mean \ac{WSS} random process, resulting in the received signal \mbox{$r(t)=s(t)\star h_\Gamma(t)+v(t)$}. Note that, due to the noise presence, $r(t)$ is also a \ac{WSS} random process.
%

At the receiver side, the signal $r(t)$ passes through an analog-to-digital converter, originating the discrete-time domain vector $\mathbf{r}\in\mathbb{R}^{(2N+L_{cp})\times1}$. Next, $\mathbf{r}$ is inputted to the function $\mathcal{Q}(\cdot)$, which synthesizes the digital processing at the receiver side of the \ac{HS-OFDM}-based \ac{TDR} system. The performed processing by this function starts with cyclic prefix removal from $\mathbf{r}$, which originates the discrete-time domain vector $\mathbf{y}\in\mathbb{R}^{2N\times1}$. 

The operations ranging from the transmitter processing on $\mathbf{x}$ to ultimately obtain $\mathbf{y}$ can be equivalently expressed in the discrete-frequency domain as
\begin{equation}\label{eq:cirD_conv_freq}
\mathbf{Y}=\mathbf{X}\odot\mathbf{H}_\Gamma+\mathbf{V},
\end{equation} 
where \mbox{$\mathbf{Y}=\frac{1}{\sqrt{2N}}\mathbf{W}_{2N}\mathbf{y}$}, $\mathbf{Y}\in\mathbb{C}^{2N\times1}$, is the discrete-frequency domain representation of $\mathbf{y}$; \mbox{$\mathbf{H}_\Gamma=\mathbf{W}_{2N}[\mathbf{h}_\Gamma^T \mathbf{0}_{1\times(2N-L_h)}]^T$}, $\mathbf{H}_\Gamma\in\mathbb{C}^{2N\times1}$, is the $2N$-length zero-padded version of the  $L_h$-length discrete-frequency equivalent of the reflection channel impulse response $h_\Gamma(t)$; and $\mathbf{V} = \frac{1}{\sqrt{2N}} \mathbf{W}_{2N} \mathbf{v}$, $\mathbf{V} \in \mathbb{R}^{2N\times1}$ is a vector composed by $2N$ proper Gaussian random variables and $\mathbf{v} \in \mathbb{R}^{2N\times1}$ is a $2N$-length discrete-time domain window of the addtive noise $v(t)$. 

Alternatively, the vector $\mathbf{V}$ can be represented as \mbox{$\mathbf{V} = [V_0, V_1,...,V_{2N-1}]^T$}. Each these variables have equal mean $\mathbb{E}\{V_k\} = 0$, $ k = 0,\cdots,2N-1$, and different variances $\mathbb{E}\{|V_k|^2\}=\sigma_{V,k}^2$, being uncorrelated in the frequency domain, i.e., \mbox{$\mathbb{E}\{V_kV_j^*\}=\mathbb{E}\{V_k\}\mathbb{E}\{V_j^*\}$} for \mbox{$k=0,\cdots,2N-1$} and \mbox{$j=0,\cdots,2N-1$} such that $k\neq j$, $k\neq 2N-j$.
Assuming that the two-sided additive noise \ac{PSD} is flat within each subband, one can represent it as the vector \mbox{$\mathbf{S_V}=[S_{V,0},S_{V,1},\cdots,S_{V,2N-1}]^T$}, \mbox{$\mathbf{S_V}\in\mathbb{R}^{2N\times1}$}, where \mbox{$S_{V,k}=\sigma_{V,k}^2/(2N\Delta f)$} is the additive noise \ac{PSD} at the $k^{th}$ subchannel and \mbox{$\Delta f=F_s/2N$} is the subcarrier frequency spacing/bandwidth.


The processing performed by $\mathcal{Q}(\cdot)$ is completed with the obtaining of an $L_\rho$-length reflectogram that is denoted in the discrete-frequency domain as $\boldsymbol{\Rho}\in\mathbb{C}^{L_\rho\times1}$, $L_\rho\geq L_h$. This is performed via processing on the discrete-frequency domain vector $\mathbf{Y}$, in such a way that the influence of the transmit vector $\mathbf{X}$ on it is minimized or eliminated. A proper reflectogram $\boldsymbol{\Rho}$ must be a good estimate of the reflection channel frequency response $\mathbf{H}_{\Gamma}$ and present a high \ac{SNR} in order for noise effect to be negligible. The processing performed by $\mathcal{Q}(\cdot)$ is finally completed with a \ac{IDFT} on $\boldsymbol{\Rho}$, resulting in the discrete-time domain reflectogram $\boldsymbol{\rho}\in\mathbb{R}^{L_\rho\times1}$ that is actually used for analyzing the reflections along the network.

After a reflectogram $\boldsymbol{\rho}$ has been obtained, analog counterpart $\rho(t)$ can be yielded in order to reduce the temporal granularity and, as a consequence, the spatial granularity of the reflectogram and the impedance discontinuity location accuracy. This can be performed via sinc interpolation in the time domain by a reconstruction filter \cite{mitra2010}. Alternatively, zero-padding can be performed on the discrete-frequency domain vector $\boldsymbol{\Rho}$ before transforming it to the time domain \cite{passerini2017}. Both these alternatives yield valid reflectograms under the condition that the signal has been sampled at sampling rate of at least $Fs=2B$. A third alternative would be oversampling $r(t)$ at a sampling frequency $Fs=2\eta B$ \cite{hassen2015}, which would ultimately yield a reflectogram $\mathbf{\rho}\in\mathbb{R}^{\eta L_\rho\times1}$, $\eta\in\mathbb{R}$, $\eta>1$, with finer temporal and spatial granularities for higher $\eta$ values. Although this is a valid alternative, it would yield significantly higher computational complexity, as all processing stages at the receiver would be performed on longer vectors.

The quality of the obtained reflectogram will ultimately depend not only on the \ac{SNR} level, but also on the processing performed on $\mathbf{y}$. Given this context, Section~\ref{sec:processing} carries out a careful description of two approaches for the obtaining of the reflectogram $\boldsymbol{\rho}$ from the received vector $\mathbf{y}$ performed by  $\mathcal{Q}(\cdot)$.
%

\section{Reflectogram Processing Approaches}\label{sec:processing}

A conventional procedure for obtaining reflectograms in \ac{TDR} systems is the pulse compression, which is based on the correlation between the transmit and received signals without cyclic prefix, i.e., $\mathbf{x}$ and $\mathbf{y}$, and has already been addressed in the literature \cite{hassen2015,hassen2018}. An alternative processing, usually resorted to in radar systems, consists of obtaining a reflectogram $\boldsymbol{\rho}$ that is an adequate estimate of $\mathbf{h}_\Gamma$ by removing the effect of the transmit signal on $\mathbf{y}$ via processing in the frequency domain \cite{sturm2009}. These two approaches are described in Subsections~\ref{subsec:pulse_compression} and \ref{subsec:channel_estimation}, respectively.

\subsection{Pulse compression}\label{subsec:pulse_compression}

The pulse compression consists of correlating the received vector $\mathbf{y}$ with a copy of the transmit vector $\mathbf{x}$ so that the obtained reflectogram $\boldsymbol{\rho}$ is equivalent to the one obtained via the transmission of a narrower pulse, which corresponds to the autocorrelation of $\mathbf{x}$ \cite{furse2006,hassen2015,parte1}.

Such procedure is performed via a linear convolution between the received $\mathbf{y}$ vector and a matched filter to $\mathbf{x}$, denoted by $\mathbf{g}\in\mathbb{R}^{2N\times 1}$. This linear convolution can be implemented by means of a circular convolution between $(4N-1)$-length zero-padded versions of $\mathbf{x}$ and $\mathbf{y}$, demanding computational complexity $\mathcal{O}((4N-1)^2)$ \cite{fitch1988,mitra2010}. Alternatively, it can be implemented in the discrete-frequency domain via multiplication of the $(4N-1)$-length zero-padded versions of $\mathbf{Y}$ and \mbox{$\mathbf{G}=\frac{1}{\sqrt{2N}}\mathbf{W}_{2N}\mathbf{g}=[X_0^*,X_1^*,\cdots,X_{2N-1}^*]^T$}, $\mathbf{G}\in\mathbb{C}^{2N\times 1}$, preceded by an \ac{DFT} and followed by an \ac{IDFT} \cite{fitch1988,hassen2018}. If \ac{FFT} and \ac{IFFT} are used, this approach becomes more efficient and faster than the circular convolution in the discrete-time domain \cite{fitch1988}, imposing the constraint that the zero-padded versions of the aforementioned vectors present length $4N$ and demanding a reduced computational complexity of \mbox{$\mathcal{O}(2(4N\log_24N)+4N)$}.

The $4N$-length zero-padded versions of $\mathbf{Y}$ and $\mathbf{G}$ are, respecively, \mbox{$\mathbf{Y}_{ZP}=\frac{1}{\sqrt{4N}}\mathbf{W}_{4N}[\mathbf{y}^T \mathbf{0}_{2N}^T]^T$}, \mbox{$\mathbf{Y}_{ZP}\in\mathbb{C}^{4N}$}, and \mbox{$\mathbf{G}_{ZP}=\frac{1}{\sqrt{4N}}\mathbf{W}_{4N}[\mathbf{g}^T \mathbf{0}_{2N}^T]^T$}, \mbox{$\mathbf{G}_{ZP}\in\mathbb{C}^{4N}$}. The resulting reflectogram can therefore be represented in the discrete-frequency domain as
\begin{eqnarray}\label{eq:pulse_compression}
	\boldsymbol{\Rho}_{PC} & = & \mathbf{Y}_{ZP}\odot\mathbf{G}_{ZP}\nonumber\\
					  & = & \mathbf{R}_{XX,ZP}\odot\mathbf{H}_{\Gamma,ZP}+\mathbf{V}_{ZP}\odot\mathbf{G}_{ZP},
\end{eqnarray} 
where $\boldsymbol{\Rho}_{PC}$ presents length $L_\rho=4N$, i.e., $\boldsymbol{\Rho}_{PC}\in\mathbb{C}^{4N\times1}$; \mbox{$\mathbf{R}_{XX,ZP}=\mathbf{X}_{ZP}\odot\mathbf{G}_{ZP}$}, \mbox{$\mathbf{R}_{XX,ZP}\in\mathbb{R}^{4N\times1}$}, represents the autocorrelation of $\mathbf{x}$ in the discrete-frequency domain; \mbox{$\mathbf{X}_{ZP}=\frac{1}{\sqrt{4N}}\mathbf{W}_{4N}[\mathbf{x}^T \mathbf{0}_{2N}^T]^T$}, \mbox{$\mathbf{X}_{ZP}\in\mathbb{C}^{4N}$}, is the $4N$-length zero-padded version of $\mathbf{X}$; \mbox{$\mathbf{H}_{\Gamma,ZP}=\mathbf{W}_{4N}[\mathbf{h}^T \mathbf{0}_{4N-L_h}^T]^T$}, \mbox{$\mathbf{H}_{\Gamma,ZP}\in\mathbb{C}^{4N}$}, is the $4N$-length zero-padded version of $\mathbf{h}$; and \mbox{$\mathbf{V}_{ZP}=\frac{1}{\sqrt{4N}}\mathbf{W}_{4N}[\mathbf{v}^T \mathbf{0}_{2N}^T]^T$}, \mbox{$\mathbf{V}_{ZP}\in\mathbb{C}^{4N}$}, is the $4N$-length zero-padded version of the $2N$-length noise discrete-time domain window $\mathbf{v}$ in the discrete-frequency domain.

A closer examination of \eqref{eq:pulse_compression} reveals that the obtained reflectogram is biased by the autocorrelation of the transmit signal $\mathbf{x}$. The resulting reconstructed analog reflectogram $\rho(t)$ obtained by the pulse-compression \ac{TDR} system will be therefore equivalent to a reflectogram obtained by a conventional \ac{TDR} system that performs injection of a pulse $s(t)=R_{xx}(t)$ into the network, being $R_{xx}(t)$ the continuous-time domain reconstructed version of the discrete-frequency domain autocorrelation of $\mathbf{x}$, i.e., \mbox{$\mathbf{R}_{XX,ZP}$}. If $R_{xx}(t)$ presents high \ac{PSLR} and \ac{ISLR} values \cite{lellouch2016,parte1}, the reflectogram $\rho(t)$ will present significant clutter and the performance of the fault detection and location procedures will therefore be compromised. Such distortion can be minimized by signal processing techniques such as windowing, which reduces the level of sidelobes at the cost of main lobe broadening and consequent resolution loss \cite{richardsvol1,temes1962,hassen2015}, or designing pulses with desired autocorrelations. The latter approach, however, may limit the use of the \ac{HS-OFDM} signal to reflectometric sensing, possibly inhibiting its use for communication purposes.

\subsection{Channel estimation}\label{subsec:channel_estimation}

In this approach, the reflectogram $\boldsymbol{\rho}$ is obtained by performing a \ac{DFT} on $\mathbf{y}$, followed by a element-wise division of the resulting vector $\mathbf{Y}$ by $\mathbf{X}$ and an \ac{IDFT} \cite{sturm2009}. The resulting operation performed by $\mathcal{Q}(\cdot)$ is therefore equivalent to a channel estimation procedure \cite{oliveira2014}.
The obtained reflectogram can be expressed in the discrete-frequency domain as
\begin{eqnarray}\label{eq:channel_estimation}
	\boldsymbol{\Rho}_{CE} & = & \mathbf{Y}\oslash\mathbf{X}\nonumber\\
					  & = & \mathbf{H}_\Gamma+\mathbf{V}\oslash\mathbf{X},
\end{eqnarray} 
with $\boldsymbol{\Rho}_{CE}$ presenting length $L_\rho=2N$, i.e., \mbox{$\boldsymbol{\Rho}_{CE}\in\mathbb{C}^{2N\times1}$}.

Finally, the discrete-time domain reflectogram $\boldsymbol{\rho}$ is then simply obtained via a \ac{IDFT} on $\boldsymbol{\Rho}$. If \ac{DFT} and \ac{IDFT} are performed by their fast counterparts, i.e., \ac{FFT} and \ac{IFFT}, the computational complexity associated with this procedure is \mbox{$\mathcal{O}(2(2N\log_2(2N))+2N)$} due to the two aforementioned stages plus the Hadamard division, being therefore significantly smaller than in the pulse compression case.

Unlike the pulse compression case, the reflectogram from \eqref{eq:channel_estimation} is unbiased. The only factors limiting the reflectogram quality will therefore be \ac{SNR} level and system bandwidth. The band limiting of the \ac{HS-OFDM}-based \ac{TDR} system to a bandwidth $B$ results in a reconstructed analog reflectogram $\rho(t)$ will be equivalent to the one obtained by the injection of $s(t)=B\text{sinc}(Bt)$, \mbox{$\text{sinc}(t)=\frac{\sin(\pi t)}{\pi t}$}, into the network by a conventional \ac{TDR} system, which is due to the resulting flat, unitary spectral content from the Hadamard division $\mathbf{X}\oslash\mathbf{X}$ that undergoes Hadamard product with $\mathbf{H}_\Gamma$ in \eqref{eq:channel_estimation}. Although clutter level may not be as relevant as in the pulse compression case, it can still be reduced by windowing \cite{sit2017}.


\section{System Parametrization and Limitations}\label{sec:limitations}

For the system described in Section~\ref{sec:model}, which provides a reflectogram via pulse compression and channel equalization approaches described in Section~\ref{sec:processing}, a proper parametrization must be performed in order for the resulting reflectogram to be representative of the network, besides achieving desired maximum unambiguous range and range resolution values. This is mainly done by adopting \ac{HS-OFDM} symbol length $2N$, cyclic prefix length $L_{cp}$, and occupied frequency bandwidth $B$ values that satisfy the aforementioned constraints, as described in the following subsections.

\subsection{Channel coherence bandwidth}\label{subsec:CBW}

In order for $h(t)$ to be appropriately reconstructed from $\mathbf{h}$, a further constraint must be satisfied besides the one related to the channel coherence time $T_c$ mentioned in Section~\ref{sec:model}. This constraint is that the frequency resolution, or subcarrier frequency spacing/bandwidth, $\Delta f$ must be smaller than the reflection channel coherence bandwidth $B_{c,H_\Gamma}$, within which the reflection channel frequency response can be considered flat. 
The reflection channel coherence bandwidth is expressed by \cite{colenSCRA}
\begin{equation}\label{eq:BcH}
B_{c,H_\Gamma} \triangleq \mathop {\max\limits_{|R_{H_\Gamma}(\Delta f)| \geq \alpha|R_{H_\Gamma}(0)|} \{\Delta f\}},
\end{equation}
where $0\leqslant\alpha\leqslant1$ is a threshold and $R_{H_\Gamma}(\Delta f)$ is the frequency correlation function expressed by
\begin{equation}
R_{H_\Gamma}(\Delta f) \triangleq \int\limits_{-\infty}^{\infty}{H_\Gamma(f)H_\Gamma^*(f+\Delta f) df}.
\end{equation}
The frequency resolution, i.e., subcarrier frequency spacing/bandwidth, must therefore be set such that $\Delta f \leq B_{c,H_\Gamma}$. As $\Delta f = F_s/(2N)$ and $Fs=2B$, it holds the constraint
\begin{equation}\label{eq:N_CBW}
	N \geq \frac{B}{B_{c,H_\Gamma}}.
\end{equation}

\subsection{Maximum unambiguous range}\label{subsec:MUR}

The maximum unambiguous range is limited by the length $2N$ of the received vector $\mathbf{y}$ \cite{nuss2018}. This is due to the fact that, in both pulse compression and channel equalization approaches, the reflectogram originates from the removal or minimization of the effect of $\mathbf{x}$ on $\mathbf{y}$. After this is done, the information on the reflection channel impulse response $\mathbf{h}$ will be contained in a $2N$-length vector. One must therefore have $2N\geq L_h$ in order for no ambiguities to occur in the reflectogram. This holds even for the pulse compression case, where the reflectogram length is only longer due to zero padding. 

Additionally, if the cyclic prefix is not long enough, there may be ambiguities in the reflectogram due to \ac{ISI} \cite{nuss2018}. In order to avoid such effect, the cyclic prefix length must satisfy $L_{cp}\geq L_h$.
Due to the aforementioned limitations, the maximum unambiguous range in both pulse compression and channel equalization approaches is
\begin{equation}\label{eq:MUR}
	d_{\max} = \frac{v_pT_s}{2}\mathop {\min \{2N,L_{cp}\}},
\end{equation}
i.e., the minimum length between $2N$ and $L_{cp}$, multiplied by the sampling period $T_s$ and the phase velocity $v_p$ and divided by $2$ in order to account for the round trip time of the reflections \cite{parte1}.
Therefore, for a given maximum unambiguous range $d_{\max}$, one has the constraints
\begin{equation}\label{eq:N_MUR}
	N \geq \frac{d_{\max}}{v_pT_s}
\end{equation}
and
\begin{equation}
	L_{cp} \geq \frac{2d_{\max}}{v_pT_s}.
\end{equation}

\subsection{Range resolution}\label{subsec:range_res}

The range resolution is the capability of resolving close reflections. For an \ac{HS-OFDM}-based \ac{TDR} system occupying a frequency bandwidth of $B$ Herz in the baseband, the corresponding range resolution is
\begin{equation}
\delta = \frac{v_p}{4B},
\end{equation}
which holds for both pulse compression \cite{hassen_tese2014,parte1} and channel equalization \cite{sturm2009} procedures.

\section{Multiple Access Schemes}\label{sec:multiple_access}

If $N_{PLM}$ \acp{PLM} are to operate over a single power distribution network, therefore constituting a distributed \ac{HS-OFDM}-based \ac{TDR} system, multiple access schemes must be adopted. The most usual ones are \ac{TDMA}, \ac{FDMA}, and \ac{CDMA} \cite{hassen_tese2014,nuss2018}.

For the pulse compression processing approach, \ac{TDMA}, \ac{FDMA}, and \ac{CDMA} schemes have already been addressed in the literature \cite{hassen_tese2014,lelong2010}. Out of this reason and due to the evident advantages of the channel estimation procedure over the pulse compression pointed out in Section~\ref{sec:processing}, the discussion on multiple access schemes for an \ac{HS-OFDM}-based \ac{TDR} system presented in the following subsections will be focused on the channel estimation procedure.

\subsection{Time-division multiple access}\label{subsec:TDMA}

\ac{TDMA} is a widely used \ac{MIMO} scheme in applications such as radar \cite{nuss2018} and communications \cite{filomeno2018}. For $N_{PLM}$ \acp{PLM} transmitting and receiving $2N$-length \ac{HS-OFDM} symbols, each one of them accessing the same power distribution network in different time slots, there will be no interference among reflectograms of the different \acp{PLM}. Furthermore, each of these reflectograms will have the same maximum unambiguous range $d_{\max,TDMA}=d_{\max}$ and range resolution $\delta_{TDMA}=\delta$ exactly as defined in Section~\ref{sec:limitations}, being also subject to the channel coherence bandwidth constraint.

In this scheme, there is a number of
\begin{equation}
	N_{\rho,TDMA}=\frac{1}{N_{PLM}T_{symb}}
\end{equation}
obtained measurements per \ac{PLM} per second. If intended, the \acp{PLM} that are not transmitting in the current time slot can also use the received signal from the transmitting \ac{PLM} for obtaining a transferogram (i.e., channel impulse response between two \acp{PLM}), which can be used for data fusion among the \acp{PLM} \cite{auzanneau2016}. This would result in a total number of \mbox{$N_{meas,TDMA}=1/T_{symb}$} measurements per \ac{PLM} per second, which is constituted by $N_{\rho,TDMA}=1/(N_{PLM}T_{symb})$ reflectograms and \mbox{$N_{t,TDMA}=(N_{PLM}-1)/(N_{PLM}T_{symb})$} transferograms.

\subsection{Frequency-division multiple access}\label{subsec:FDMA}

In the case where different \acp{PLM} are to transmit simultaneously, an alternative is the use of the \ac{FDMA} principle. For better exploiting information on the network, an interleaved \ac{FDMA} scheme can be used rather than a localized one that allocated adjacent subcarriers to each \ac{PLM} \cite{sturm2013,hassen2015}. In this approach, $2N/N_{PLM}$ interleaved subcarriers are allocated to each \ac{PLM} in an \ac{HS-OFDM}-based \ac{TDR} system with \ac{HS-OFDM} symbol length $2N$. The resulting reflectogram for each \ac{PLM} is therefore obtained by discarding the subcarriers allocated to the remaining \acp{PLM} and, after obtaining the discrete-frequency domain reflectogram $\boldsymbol{\Rho}$, performing a $(2N/N_{PLM})$-size \ac{IDFT} on it in order to obtain the discrete-time domain reflectogram $\boldsymbol{\rho}$. Finally, the obtained $(2N/N_{PLM})$-length reflectogram $\boldsymbol{\rho}$ undergoes a reconstruction process that will originate $\rho(t)$ as described in Section~\ref{sec:model}.

In the considered \ac{HS-OFDMA} scheme, the vector \mbox{$\mathbf{D}=[D_0,D_1,\cdots,D_{N-1}]^T$}, \mbox{$\mathbf{D}\in\mathbb{C}^{N\times1}$}, from Section~\ref{sec:model} is generated via a pre-mapping on a vector \mbox{$\dot{\mathbf{D}}=[\dot{D}_0,\dot{D}_1,\cdots,\dot{D}_{N}]^T$}, $\dot{\mathbf{D}}\in\mathbb{C}^{(N+1)\times1}$, whose $k^{th}$ sample $\dot{D}_{k}$ will be ultimately transmitted at the continuous-frequency bin \mbox{$f_k=k\Delta f$} Hz after all processing and digital-to-analog conversion has been performed at the transmitter side. This pre-mapping is expressed as
\begin{equation}\label{eq:premap_OFDMA}
D_k = \left\{\arraycolsep=3pt\def\arraystretch{1.3}
\begin{array}{ll}
\dot{D}_{k+1}, & k=0,\cdots,N-2\\
\dot{D}_{0}+j\dot{D}_{N}, & k=N-1
\end{array}\right. ,
\end{equation}
where $j=\sqrt{-1}$. In order for interference-free operation of the \acp{PLM} to take place, a set \mbox{$\mathcal{K}_u=\{u,u+N_{PLM},u+2N_{PLM},\cdots\}$}, \mbox{$u=0,\cdots,N_{PLM}-1$}, of $N/N_{PLM}$ interleaved samples of $\dot{\mathbf{D}}$ are allocated to the $u^{th}$ \ac{PLM}. Also, all \acp{PLM} place null values at the samples that have not been allocated to it.

Once the discrete-frequency domain reflectogram $\boldsymbol{\Rho}$ has been obtained at the receiver side, the \ac{PLM} performs a channel estimation processing on the corresponding samples to the frequency bins allocated to it. Next, $\boldsymbol{\Rho}$ is transformed into a modified reflectogram according to the mapping
\begin{equation}\label{eq:map_OFDMA}
\dot{\Rho}_k = \left\{\arraycolsep=3pt\def\arraystretch{1.3}
\begin{array}{ll}
\Rho_k, & k=0,N\\
2\Rho_k, & k=1, ..., N-1
\end{array}\right. ,
\end{equation}
in which $\dot{\Rho}_k$ is the $k^{th}$ sample of the modified reflectogram \mbox{$\dot{\boldsymbol{\Rho}}=[\dot{\Rho}_0,\dot{\Rho}_1,\cdots,\dot{\Rho}_{N}]^T$}, \mbox{$\dot{\boldsymbol{\Rho}}\in\mathbb{R}^{(N+1)\times1}$}. Such mapping is performed in order to map a $2N$-length Hermitian symmertric vector $\boldsymbol{\Rho}$ into the $N$-length vector $\dot{\boldsymbol{\Rho}}$ with no loss of spectral content.

Finally, the $u^{th}$ \ac{PLM} obtains an $L_\rho$-length discrete-time domain reflectogram \mbox{$\boldsymbol{\rho}_u=[\rho_{u,0},\rho_{u,1},\cdots,\rho_{u,L_\rho}]^T$}, \mbox{$\boldsymbol{\rho}_u\in\mathbb{R}^{L_\rho\times1}$}, \mbox{$L_\rho=2N/N_{PLM}$}, whose $n^{th}$ sample is expressed as
\begin{eqnarray}\label{eq:reflectogram_OFDMA}
	\rho_{u,n} & = & \frac{1}{\sqrt{L_\rho}}\Bigg[\sum_{k\in\mathcal{K}_u}\Re\{\dot{\Rho}_k\}\cos\left(\frac{\pi k n}{N}\right)\nonumber\\
	& - &  \sum_{k\in\mathcal{K}_u}\Im\{\dot{\Rho}_k\}\sin\left(\frac{\pi k n}{N}\right)\Bigg],
\end{eqnarray}
which, for the single-\ac{PLM} case, is exactly the same result obtained by performing an \ac{IDFT} on $\boldsymbol{\Rho}$.

The obtained reflectograms in the proposed \ac{HS-OFDMA} scheme will have maximum unambiguous range experienced by each \ac{PLM} equal to \mbox{$d_{\max,FDMA}=d_{\max}/N_{PLM}$}. The range resolution for each \ac{PLM} in its turn will be \mbox{$\delta_{FDMA}=v_p/(4B_{FDMA})$}, with $B_{FDMA}$ being the effective bandwidth experienced by each \ac{PLM}. As $2N/N_{PLM}$ subcarriers are allocated to each \ac{PLM}, the effective subcarrier separation experienced by the latter will be \mbox{$\Delta f_{FDMA} = N_{PLM}B/N$}. Consequently, $B_{FDMA}=N \Delta f_{FDMA}/N_{PLM}=B$, and it holds that the range resolution is exactly as defined in Section~\ref{sec:limitations} for the single-user case, i.e., $\delta_{FDMA}=\delta$. 
Although the range resolution is maintained, the information on the network contained in the subcarriers discarded by the \ac{PLM} will cause distortion on the obtained reflectograms. The trade-off between number of \acp{PLM} connected to the network and reflectogram quality must therefore be observed in order for fault detection and location procedures not to be negatively affected.

As a consequence, one obtains a number of
\begin{equation}
N_{\rho,FDMA}=\frac{1}{T_{symb}}
\end{equation}
reflectograms per \ac{PLM} per second. If transferograms are also to be obtained, one would have a total number of \mbox{$N_{meas,FDMA}=N_{PLM}/T_{symb}$} measurements per \ac{PLM} per second, encompassing \mbox{$N_{\rho,FDMA}=1/T_{symb}$} reflectograms and \mbox{$N_{t,FDMA}=(N_{PLM}-1)/T_{symb}$} transferograms.

%

\subsection{Code-division multiple access}\label{subsec:CDMA}

Unlike the \ac{TDMA} and \ac{FDMA} schemes, the use of \ac{CDMA} allows the \acp{PLM} to occupy the entire bandwidth over all time slots \cite{lelong2010,hassen_tese2014,nuss2018}. The implementation of this scheme consists of adding an encoding process on the discrete-frequency domain vector $\mathbf{D}$.

In this context, let us consider an $N_{PLM}$-size Hadamard matrix whose rows are constituted by codewords that are orthogonal among each other. The codeword of the $u^{th}$ row of this matrix, $u=0,\cdots,N_{PLM}-1$, is represented by the vector \mbox{$\mathbf{C}_u=[C_{u,0},C_{u,1},\cdots,C_{u,N_{PLM}-1}]^T$}, \mbox{$\mathbf{C}_u\in\{-1,1\}^{N_{PLM}\times1}$}, which is in its turn allocated to the $u^{th}$ \ac{PLM}. Due to the orthogonality among the codewords, the mutual interference among the \acp{PLM} is minimized.

The encoding process performed by the $u^{th}$ \ac{PLM} is then performed via a multiplication between the codeword vector by the transpose counterpart of $\mathbf{D}\in\mathbb{C}^{N\times1}$, resulting in the spread symbol matrix
\begin{equation}
	\mathbf{D}_{C,u} = \mathbf{C}_u\mathbf{D}^T,
\end{equation}
for which holds $\mathbf{D}_{C,u}\in\mathbb{C}^{N_{PLM}\times N}$. The $N_{PLM}$ rows of this matrix are then sequentially fed to the Hermitian symmetric mapping and further processing, being finally transmitted by the $u^{th}$ \ac{PLM}. Thus, the resulting transmission time is $N_{PLM}$ times longer than the $T_{symb}$ seconds of the single-\ac{PLM} case.

After cyclic prefix removal and \ac{DFT} are performed on the $N_{PLM}$ received symbols at the receiver side, a reception matrix $\mathbf{Y}_{C}\in\mathbb{C}^{N_{PLM}\times2N}$ is formed, being its rows associated with the rows of $\mathbf{D}_{C,u}$. 
\begin{equation}\label{eq:recep_matrix}
\mathbf{Y}_{C} = \mathbf{X}_{C}\odot\mathbf{H}_{\Gamma,C} + \mathbf{V}_{C} + \mathbf{V}_{C,I},
\end{equation}
where $\mathbf{X}_{C}\in\mathbb{C}^{N_{PLM}\times2N}$ is a transmission matrix whose rows result from the Hermitian symmetric mapping on the respective rows of $\mathbf{D}_{C,u}$; $\mathbf{H}_{\Gamma,C}\in\mathbb{C}^{N_{PLM}\times2N}$ is an equivalent reflection channel matrix, expressed as $\mathbf{H}_{\Gamma,C}=\mathbf{1}_{N_{PLM}\times1}\mathbf{H}_{\Gamma}^T$; $\mathbf{V}\in\mathbb{C}^{N_{PLM}\times2N}$ is a noise matrix whose rows are $2N$-length discrete-frequency domain windows of the additive noise $v(t)$; and $\mathbf{V}_{I}$ is the interference noise matrix, which accounts for the mutual interference among the \acp{PLM}. The $u^{th}$ \ac{PLM} then performs a decoding process on $\mathbf{Y}_{C}$, which generates the vector $\mathbf{Y}\in\mathbb{C}^{2N\times1}$ and is expressed as
\begin{equation}\label{eq:decod}
	\mathbf{Y} = \frac{\left(\mathbf{C}_u^T\mathbf{Y}_{C}\right)^T}{\left|\mathbf{C}_u\right|},
\end{equation}
where $\left|\mathbf{C}_u\right|=\mathbf{C}_u^T\mathbf{C}_u$ is the cardinality of the codeword vector $\mathbf{C}_u$. The discrete-frequency domain vector $\mathbf{Y}$ can be alternatively expressed as
\begin{equation}\label{eq:decod2}
\mathbf{Y} = \mathbf{X}\odot\mathbf{H}_\Gamma + \hat{\mathbf{V}} + \hat{\mathbf{V}}_I,
\end{equation}
in which $\hat{\mathbf{V}}=\left(\mathbf{C}_u^T\mathbf{V}_{C}\right)^T/\left|\mathbf{C}_u\right|$ is the resulting additive noise from the decoding process, whose \ac{PSD} is expressed as \mbox{$\mathbf{S_{\hat{V}}}=[S_{\hat{V},0},S_{\hat{V},1},\cdots,S_{\hat{V},2N-1}]^T$}, \mbox{$\mathbf{S_{\hat{V}}}\in\mathbb{R}^{2N\times1}$}, with \mbox{$S_{\hat{V},k}=(\sigma_{V,k}^2/\sqrt{N_{PLM}})/(2N\Delta f)$} representing the resulting additive noise \ac{PSD} at the $k^{th}$ subchannel; and \mbox{$\hat{\mathbf{V}}_I=\left(\mathbf{C}_u^T\mathbf{V}_{C,I}\right)^T/\left|\mathbf{C}_u\right|$} is the resulting interference noise. The fact that the resulting additive noise has variance and therefore \ac{PSD} $\sqrt{N_{PLM}}$ times smaller than the ones of the original additive noise $\mathbf{V}$ is due to the averaging process that happens at the decoding stage of the \ac{CDMA} scheme in \eqref{eq:decod} and ultimately results in an \ac{SNR} gain \cite{lelong2010}.

After this stage, a channel estimation procedure is performed on $\mathbf{Y}$ as described in Subsection~\ref{subsec:channel_estimation} for the single-\ac{PLM} case. The resulting discrete-frequency domain reflectogram $\boldsymbol{\Rho}$ finally undergoes an \ac{IDFT}, originating the discrete-time domain reflectogram $\boldsymbol{\rho}$. Although the effect of the transmit symbol on the reflectogram is properly removed, the use of \ac{CDMA} introduces mutual interference among the \acp{PLM}, whose reflectograms consist of the reflections raised by the injected signal by the own \ac{PLM} plus a term that accounts for the decoded symbols transmitted by the remaining \acp{PLM} as well as additive noise. Despite being significantly attenuated by the decoding process, this interference term might distort the obtained reflectogram \cite{nuss2018}.

In spite of the undesired interference, the obtained reflectograms in the \ac{CDMA} scheme yield the same maximum unambiguous range $d_{\max,CDMA}=d_{\max}$ and range resolution $\delta_{CDMA}=\delta$ as in the single-user case from Section~\ref{sec:limitations}, whose channel coherence bandwidth constraint still holds. Due to the extended transmission time, a number
\begin{equation}
N_{\rho,CDMA}=\frac{1}{N_{PLM}T_{symb}}
\end{equation}
of reflectograms are obtained per \ac{PLM} per second. In the case where transferograms are also obtained, the total number of measurements per \ac{PLM} per second would be \mbox{$N_{meas,CDMA}=1/T_{symb}$} measurements, which encompasses \mbox{$N_{\rho,CDMA}=1/(N_{PLM}T_{symb})$} reflectograms and \mbox{$N_{t,CDMA}=(N_{PLM}-1)/(N_{PLM}T_{symb})$} transferograms.
%

\section{Numerical Analysis}\label{sec:analysis}

In this section, the carried out discussion on reflectogram processing approaches and multiple access schemes throughout the paper is validated via numerical results. Given this context, Subsection~\ref{subsec:comparison} compares the pulse compression and channel estimation procedures, while Subsection~\ref{subsec:limitations} addresses system limitations due to regulatory constraints. Finally, the discussed multiple access schemes are analyzed in Subsection~\ref{subsec:MA}.

\subsection{Pulse compression and channel estimation comparison}\label{subsec:comparison}

\begin{figure}[!t]
	\centering
	\psfrag{XX}[c][c]{$N$}
	\psfrag{YY}[c][c]{$\mathcal{O}$}
	\includegraphics[width=8.5cm]{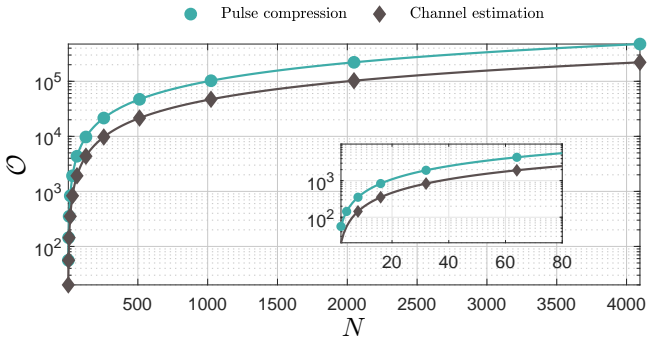}
	\caption{Computational complexity (disregarding signal reconstruction) as a function of $N$ for the two reflectogram processing approaches.}\label{fig:CC}
\end{figure}

In order to corroborate with the claim that the channel estimation procedure outperforms the classical pulse compression for \ac{TDR}, Fig.~\ref{fig:CC} presents a comparison of both approaches in terms of computational complexity. Based on the expressions provided in Subsections~\ref{subsec:pulse_compression} and \ref{subsec:channel_estimation}, a far higher computational complexity is demanded by the pulse compression approach as it is performed on a longer vector.

\begin{figure}[!t]
	\centering
	\psfrag{XX}[c][c]{$N$}
	\psfrag{YY}[c][c]{PSLR~(dB)}
	\includegraphics[width=8.5cm]{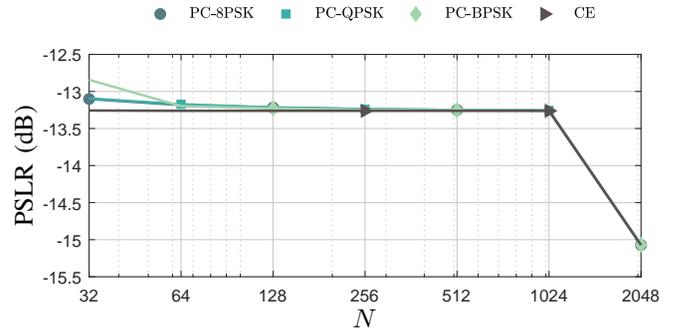}
	\caption{PSLR as a function of the number of subcarriers $N$ for the pulse compression procedure with BPSK, QPSK, and 8PSK modulations, and for the channel estimation procedure.}\label{fig:PSLR_N}
	\vspace{0.5cm}
\end{figure}

\begin{figure}[!t]
	\centering
	\psfrag{XX}[c][c]{$N$}
	\psfrag{YY}[c][c]{ISLR~(dB)}
	\includegraphics[width=8.5cm]{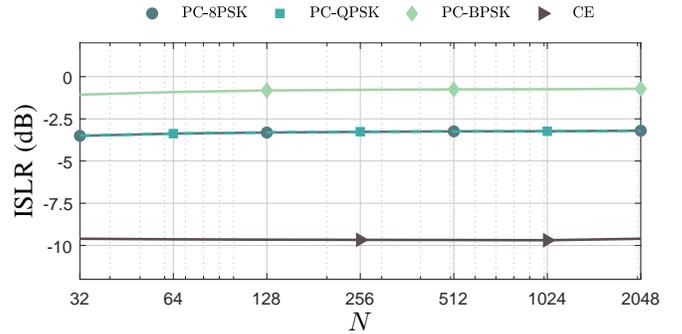}
	\caption{ISLR as a function of the number of subcarriers $N$ for the pulse compression procedure with BPSK, QPSK, and 8PSK modulations, and for the channel estimation procedure.}\label{fig:ISLR_N}
\end{figure}

Furthermore, the claim that the quality of obtained reflectograms is higher in the channel estimation procedure is endorsed by the results from Figs.~\ref{fig:PSLR_N} and \ref{fig:ISLR_N}. In these figures, \ac{PSLR} and \ac{ISLR} values are respectively shown for the equivalent transmit pulse $s(t)=R_{xx}(t)$ obtained via pulse compression and the equivalent transmit pulse $s(t)=Bsinc(Bt)$ for the channel estimation procedure. In this figure, typical modulations schemes for \ac{PLC} systems, namely \ac{BPSK}, \ac{QPSK}, and \ac{8PSK} \cite{1901_2}, are considered for the pulse compression approach, being the results for the channel estimation approach independent of the adopted scheme. 

The attained results show that the \ac{PSLR} decreases with the modulation order for the pulse compression approach. As $N$ increases, the difference among the \ac{PSLR} values yielded by either pulse compression or channel estimation becomes negligible. Finally, it is observed that \ac{PSLR} levels decrease inversely along with $N$. 

Regarding \ac{ISLR}, the attained values also decrease with the modulation order for the pulse compression procedure, being the difference between \ac{QPSK} and \ac{8PSK} more subtle. The best \ac{ISLR} values are attained by the channel estimation procedures, which significantly outperforms the pulse compression procedure by about $6.4$~dB if \ac{QPSK} and \ac{8PSK} are assumed for the latter, and by $8.9$~dB if pulse compression is performed on a signal belonging to a \ac{BPSK} modulation. Unlike the \ac{PSLR} case, there is a slight increase of \ac{ISLR} values along with $N$. This is explained by the fact that, given a fixed $B$, the side lobe level is increased, while the main lobe duration of $1/(2B)$ seconds and therefore the main lobe level are maintained for longer \ac{HS-OFDM} symbols.

\subsection{System limitations}\label{subsec:limitations}

Although an adequate parametrization of the \ac{HS-OFDM}-based \ac{TDR} system can yield valid reflectograms, regulatory constraints limit the \ac{PSLR} and \ac{ISLR} values, as well as achievable range resolution and maximum unambiguous range. The latter two parameters are also influenced by characteristics of the cable where the injected signal propagates, which define the phase velocity $v_p$. An awareness of such limitations is therefore paramount for performing adequate sensing of the power distribution network.

In this context, two scenarios are addressed, namely an European underground low-voltage power distribution network and an US overhead medium-voltage power distribution network in a rural area. For the low-voltage scenario, it is considered a power supply cable NAYY150SE with resistance $R'$, inductance $L'$, conductance $G'$, and capacitance $C'$ per unit length calculated as in \cite{lampe_vinck2011}, whereas for the medium-voltage cable, the power supply cable with distributed parameters listed in \cite{1901_2} is adopted.  Based on these parameters, the phase velocity is calculated by $v_p=1/\sqrt{L'C'}$ \cite{paul2007}, resulting in $v_p=1.50\times10^8$ for the considered \ac{LV} cable, and $v_p=2.56\times10^8$ for the considered \ac{MV} cable.

\begin{figure}[!t]
	\centering
	\psfrag{XX}[c][c]{$B$(kHz)}
	\psfrag{YY}[c][c]{$\delta$(m)}
	\includegraphics[width=8.5cm]{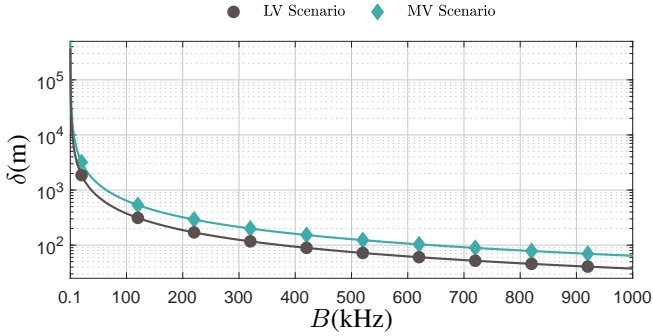}
	\caption{Range resolution $\delta$ in meters as a function of the occupied frequency bandwidth $B$ in the considered LV and MV scenarios.}\label{fig:RR}
\end{figure}

Fig.~\ref{fig:RR} shows the range resolution $\delta$ as a function of the occupied frequency bandwidth $B$ for the considered \ac{LV} and \ac{MV} scenarios.  The achieved $\delta$ values range from tens of thousands of kilometers for low $B$ values to a few meters for higher $B$ values, with a ratio of $1.71$ between the resolution in the \ac{MV} and \ac{LV} and  scenarios due to their different phase velocity. The presented results indicate that $B$ values in the \ac{NB}-\ac{PLC} frequency range, i.e., $B<500$~kHz, result in fair range resolution values, i.e., $\delta\geq75$~m, and therefore a fair capability of resolving close impedance discontinuities for typical distances covered by \ac{PLC} signaling in \ac{LV} and \ac{MV} power distribution networks, which are about $1$~km in \ac{MV} scenarios and shorter in \ac{LV} scenarios \cite{passerini2018_2,PLCbook2016}.

\begin{figure}[!t]
	\centering
	\psfrag{XX}[c][c]{$L_{cp}$}
	\psfrag{YY}[c][c]{$d_{\max}$(m)}
	\hspace*{-4mm}
	\includegraphics[width=9.75cm]{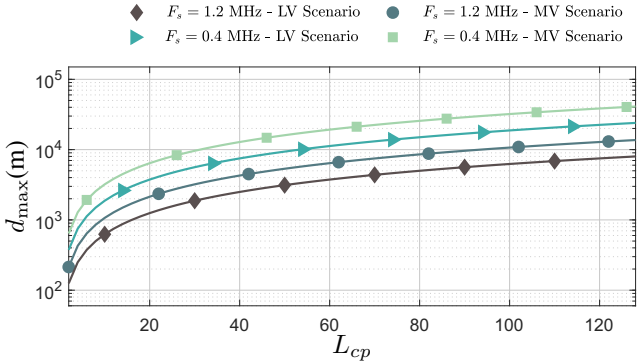}
	\caption{Maximum unambiguous range $d_{\max}$ as a function of cyclic prefix length $L_{cp}$ for a sampling frequency \mbox{$F_s=1.2$~MHz}.}\label{fig:MUR}
\end{figure}

For typical \ac{PLC} systems, one observes $2N>L_{cp}$. As a consequence, only $L_{cp}$ limits $d_{\max}$ according to the relation in \eqref{eq:MUR}. Based on this assumption and considering typical sampling frequency values for \ac{NB}-\ac{PLC}, namely $F_s=0.4$~MHz and $F_s=1.2$~MHz, Fig.~\ref{fig:MUR} shows the maximum unambiguous range $d_{\max}$ as a function of the cyclic prefix length in the considered \ac{LV} and \ac{MV} scenarios. In this figure, one observes a minimum $d_{\max}$ value of about $1$~km for $L_{cp}\geq16$, which is experienced with $F_s=1.2$~MHz in the \ac{LV} scenario. For $L_{cp}\geq64$, it holds $d_{\max}\geq4$~km for both considered sampling frequencies in both \ac{LV} and \ac{MV} scenarios. As the typical spacing between \acp{PLM} in a power distribution network is considerably shorter than that, these results show that typical cyclic prefix lengths for \ac{NB}-\ac{PLC} systems, i.e., $L_{cp}=32$ and $L_{cp}=50$, are enough for providing fair maximum unambiguous range values.

Focusing on regulatory constraints, the considered regulations in this paper are \ac{FCC}, \ac{ARIB}, and \ac{CENELEC}, which are addressed in the IEEE 1901.2 Standard \cite{1901_2} for \ac{NB}-\ac{PLC}. The frequency range, occupied frequency bandwidth $B$, sampling frequency $F_s$, \ac{FFT}/\ac{IFFT} size, range and number of active subcarriers $N_\text{active}$, and cyclic prefix length $L_{cp}$ associated with these regulations are listed in Table~\ref{tab:reg_parameters}. In this table, the a number of ative subcarriers $N_\text{active}$ is adopted for both \ac{FCC} and \ac{ARIB} regulations for covering the whole frequency band covered by these regulations described in \cite{1901_2}. Also, the whole \ac{CENELEC} band is considered, i.e., from the lowest frequency of the \ac{CENELEC} A band to the highest frequency of the \ac{CENELEC} D band, being the adopted number of active subcarriers equal to the necessary for covering this whole frequency bandwidth. For \ac{HS-OFDM}-based \ac{TDR} systems parametrized according to these three regulations, Table~\ref{tab:reg_results} lists the achieved \ac{PSLR} and \ac{ISLR} values, as well as range resolution and maximum unambiguous range for the considered \ac{LV} and \ac{MV} scenarios. Based on the data of this table, one can conclude that \ac{FCC} and \ac{ARIB} regulations allow a more appropriate sensing of shorter power distribution network sections than \ac{CENELEC}, providing finer range resolution and shorter maximum unambiguous range. Also, adopting the long $L_{cp}$ of both \ac{FCC} and \ac{ARIB} regulations results in a significant increase of $d_{\max}$ at the cost of obtaining less reflectograms over time due to the transmission of more cyclic prefix samples.

\begin{table}[!b]
	\renewcommand{\arraystretch}{1.5}
	\arrayrulecolor[HTML]{708090}
	\setlength{\arrayrulewidth}{.1mm}
	\setlength{\tabcolsep}{4pt}
	
	\footnotesize
	\centering
	\caption{Adopted \ac{HS-OFDM} system parameters according to \ac{NB}-\ac{PLC} regulations.}
	\label{tab:reg_parameters}
	\begin{tabular}{c|c|c|ccc}
		\hline\hline
		\multicolumn{3}{c|}{Regulation} 		   & FCC 	   & ARIB      & CENELEC \\ \hline
		\multicolumn{3}{c|}{Frequency range (kHz)} & $10-490$  & $10-450$  & $3-148.5$ \\ \hline
		\multicolumn{3}{c|}{$B$~(kHz)}        	   & $480$     & $440$     & $145.5$ \\ \hline
		\multicolumn{3}{c|}{$F_s$~(MHz)}      	   & $1.2$     & $1.2$     & $0.4$ \\ \hline
		\multicolumn{3}{c|}{FFT/IFFT size}         & $256$     & $256$     & $256$ \\ \hline
		\multicolumn{3}{c|}{Active subcarriers}    & $3-104$   & $3-96$    & $2-95$ \\ \hline
		\multicolumn{3}{c|}{$N_\text{active}$}    & $102$     & $94$      & $94$ \\ \hline
		\multirow{2}{*}{$L_{cp}$}   & \multicolumn{2}{c|}{Standard}             
		& $30$ 	& $30$  & $30$     \\
		& \multicolumn{2}{c|}{Long}             & $52$ 	& $52$  & -     \\ \hline\hline
	\end{tabular}
\end{table}

\subsection{Channel estimation with multiple access}\label{subsec:MA}

For a comparison among the three multiple access schemes described in Section~\ref{sec:multiple_access} that allow the simultaneous obtaining of reflectograms by multiple \acp{PLM} connected to the same power distribution network, an \ac{HS-OFDM}-based \ac{TDR} system parametrized according to the \ac{FCC} regulation is considered in this subsection. Additionally, \ac{BPSK} modulation scheme was adopted.

\begin{figure*}[!t]
	\centering	
	\psfrag{A}[c][c]{\footnotesize $d_a=1$~km}
	\psfrag{B}[c][c]{\footnotesize $d_b=1.73$~km}
	\psfrag{T}[c][c]{MV}
	\includegraphics[height=4.5cm]{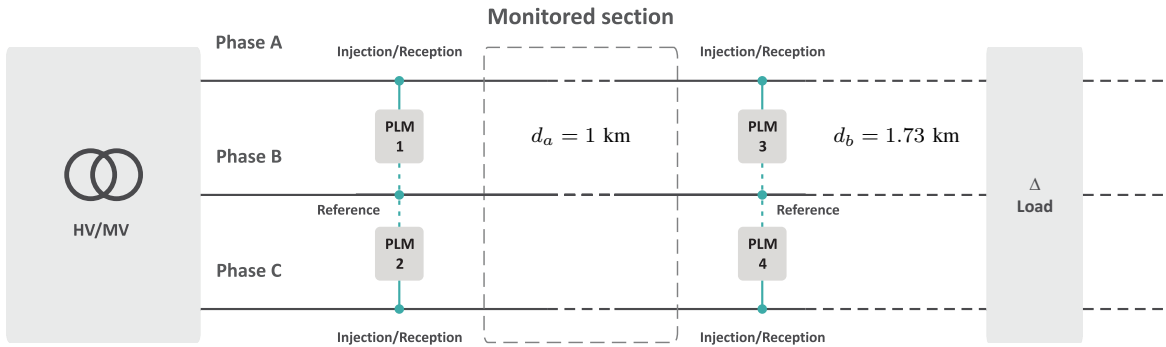}
	\vspace{0.5cm}
	\caption{Considered MV distribution network section.}\label{fig:MV_line}	
\end{figure*}

In this context, simulations were carried out based on an \ac{MTL}-based model \cite{paul2007,franek2017} of a real \ac{MV} power distribution network section in the city of Curitiba, Brazil, depicted in Fig.~\ref{fig:MV_line}. This section is consisted by a feeder, which is an \ac{HV}/\ac{MV} transformer, followed by $2.73$~km of line and a delta load. The input impedance of the \ac{HV}/\ac{MV} transformer at the \ac{MV} side, as well as the input impedance of the network after the delta load are assumed to be much higher than the characteristic impedance of the line, thus behaving as an open circuit. Two \acp{PLM} are connected to this network right after the feeder, being one connected between phases A and B and the other between phases C and B. After an $1$~km line section, other \acp{PLM} are connected to the network between the same pairs of phases. Thus, there is a line section with length of $1.73$~km between the latter \acp{PLM} and the delta load.
Regarding the additive noise, the adopted model is the one reported in \cite{tao2007,girotto2017}, which presents one-sided \ac{PSD} in the continuous-frequency domain equal to
$S_V(f)=-93+52.98e^{-0.0032f/10^3}$~dBm/Hz.

\begin{table}[!b]
	\renewcommand{\arraystretch}{1.5}
	\arrayrulecolor[HTML]{708090}
	\setlength{\arrayrulewidth}{.1mm}
	\setlength{\tabcolsep}{4pt}
	
	\footnotesize
	\centering
	\caption{Resulting range resolution in \ac{LV} and \ac{MV} scenarios, \ac{PSLR}, and \ac{ISLR} for \ac{NB}-\ac{PLC} regulations.}
	\label{tab:reg_results}
	\begin{tabular}{c|c|c|ccc}
		\hline\hline
		\multicolumn{3}{c|}{Regulation}                                   & FCC & ARIB & CENELEC \\ \hline
		\multirow{2}{*}{$\delta$~(m)}   & \multicolumn{2}{c|}{LV}             & $78.07$ & $85.17$  & $257.56$     \\
		& \multicolumn{2}{c|}{MV}             & $133.59$ & $145.73$  & $440.70$     \\ \hline
		\multirow{4}{*}{$d_{\max}$~(km)} & \multirow{2}{*}{Std. $L_{cp}$} & LV & $1.87$ & $1.87$  & $5.62$     \\
		&                                & MV & $3.21$ & $3.21$  & $9.62$     \\ \cline{2-6} 
		& \multirow{2}{*}{Long $L_{cp}$} & LV & $3.25$ & $3.25$  & -     \\
		&                                & MV & $5.56$ & $5.56$  & -     \\ \hline
		\multirow{2}{*}{Sidelobe level}   & \multicolumn{2}{c|}{PSLR (dB)}             & $-13.26$ & $-13.26$  & $-13.26$     \\
		& \multicolumn{2}{c|}{ISLR (dB)}             & $-9.66$ & $-9.66$  & $-9.66$     \\ \hline\hline
	\end{tabular}
\end{table}

%

\begin{figure}[!b]
	\centering
	\psfrag{XX}[c][c]{$N_{PLM}$}
	\psfrag{YY}[c][c]{$N_{\rho}\left(\text{s}^{-1}\right)$}
	\hspace*{-4mm}
	\includegraphics[width=9.75cm]{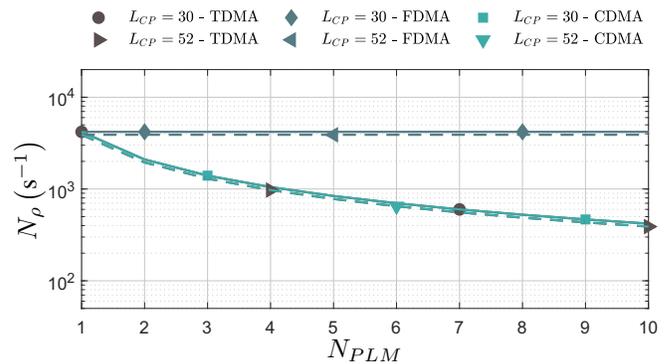}
	\caption{Number of obtained reflectograms $N_{\rho}$ as a function of the number of \acp{PLM} $N_{PLM}$ for the considered multiple access schemes and cyclic prefix lengths.}\label{fig:N_rho}
\end{figure}

The number of reflectograms obtained over time $N_\rho$, which varies due to the different procedures adopted by the multiple access schemes for providing orthogonality among the signals of the multiple \acp{PLM}, is shown in Fig.~\ref{fig:N_rho} as a function of the number of \acp{PLM} $N_{PLM}$ for both standard and short cyclic prefix lengths. In this figure, one observes decreasing $N_\rho$ values along with $N_{PLM}$ for both \ac{TDMA} and \ac{CDMA} schemes, while the \ac{FDMA} scheme presents constant $N_\rho$ regardless of the number $N_{PLM}$ of \acp{PLM}. This happens due to the long time interval during which a \ac{PLM} does not transmit signals in the \ac{TDMA} scheme. The lower $N_\rho$ values in the \ac{CDMA} scheme are on the other hand due to the spreading of \ac{HS-OFDM} symbols, which results in longer effective transmission time for \ac{HS-OFDM} symbols in this scheme. It is worth highlighting that, for a given $N_{PLM}$, the number of obtained reflectograms $N_\rho$ is the same for both \ac{TDMA} and \ac{CDMA} schemes.

As previously mentioned, the use of \ac{FDMA} results in a fixed $N_\rho$. This is shown in Fig.~\ref{fig:N_rho} based on the assumption that $N\geq N_\rho$, i.e., the number of subcarriers is not less than the number of \acp{PLM}. Nevertheless, the obtained reflectograms by each \ac{PLM} will be distorted if the number of subcarriers $N/N_{PLM}$ ($128/N_{PLM}$ in the \ac{FCC} case) assigned to each of them does not satisfy both the reflection channel coherence bandwidth constraint from \eqref{eq:N_CBW} and the maximum unambiguous range constraint from \eqref{eq:N_MUR}, being the latter only relevant if $2N/N_{PLM}<L_{cp}$ due to the relationship from \eqref{eq:MUR}.

A further analysis is a comparison of the \ac{TDMA}, \ac{FDMA}, and \ac{CDMA} multiple access schemes considering an one-sided transmit signal \ac{PSD} of $-36.81$~dBm/Hz. Consequently, the total transmission power allocated to all active subcarriers of each \acp{PLM} considering the \ac{FDMA} scheme is therefore $13.98$~dBm, while a total of $20$~dBm are allocated to each \ac{PLM} in the \ac{TDMA} and \ac{CDMA} schemes. In order to assess the performance of the aforementioned schemes, the \ac{SINR} of the discrete-frequency domain received vector $\mathbf{Y}$ was adopted as a metric. For \ac{TDMA} and \ac{FDMA} schemes, the \ac{SINR} for each \acp{PLM} is equal to the ratio between the total received signal power and the total additive noise power, being equivalent the overall \ac{SNR} considering the allocated subcarriers to them. The \ac{CDMA} case, on the other hand, has also an interference noise term resulting from the decoding process from \eqref{eq:decod} and pointed out in \eqref{eq:decod2}. As a consequence, the \ac{SINR} at the four \acp{PLM} is equal to the ratio between their associated total received signal power and the total additive noise plus interference noise power.

The attained \ac{SINR} values are listed in Table~\ref{tab:SINR}. The lower \ac{SINR} experienced by the \acp{PLM} in the \ac{TDMA} scheme is due to the fact that the whole additive noise spectral content impairs the captured signal from the reflection channel. Meanwhile, the subcarrier hopping in the \ac{FDMA} scheme results in higher \ac{SINR}. This is due to the fact that greater part of the additive noise power is concentrated in lower frequencies as its \ac{PSD} decreases exponentially. As a consequence, the subcarrier hopping makes the \acp{PLM} experience less effective noise power and therefore increases the \ac{SINR} in comparison to the \ac{TDMA} scheme. The increasing \ac{SINR} along with the \ac{PLM} index $u$ is due to the fact that subcarriers belonging to the set $\mathcal{K}_u$ allocated to the $u^{th}$ \ac{PLM} are associated to higher frequency bins $f_k=k\Delta f$ as $u$ increases, which is described in Subsection~\ref{subsec:FDMA} and results in lower effective additive noise power for higher $u$. Finally, the higher average \ac{SINR} among the \acp{PLM} in the \ac{CDMA} scheme is due to the \ac{SNR} gain due to noise averaging in the decoding process, as described in Subsection~\ref{subsec:CDMA}.


\begin{table}[!t]
	\renewcommand{\arraystretch}{1.5}
	\arrayrulecolor[HTML]{708090}
	\setlength{\arrayrulewidth}{.1mm}
	\setlength{\tabcolsep}{4pt}
	
	\footnotesize
	\centering
	\caption{\ac{SINR} at the four \acp{PLM} for the considered multiple access schemes.}
	\label{tab:SINR}
	\begin{tabular}{c|cccc}
		\hline\hline
		PLM index $u$  & 1       & 2       & 3       & 4       \\ \hline
		TDMA & $28.1595$dB & $28.1581$dB & $28.1595$dB & $28.1581$dB \\ \hline
		FDMA & $29.3370$dB & $29.8451$dB & $31.9130$dB & $32.6272$dB \\ \hline
		CDMA & $31.1698$dB & $31.1684$dB & $31.1698$dB & $31.1684$dB \\ \hline\hline
	\end{tabular}
\end{table}

\section{Conclusion}\label{sec:conclusion}

This study has discussed the main aspects of an \ac{HS-OFDM}-based \ac{TDR} system for power line sensing. In summary, a system model covering the injection of  \ac{HS-OFDM} signals into the power distribution grid and subsequent reception of raised reflections has been outlined and pulse compression and channel estimation approaches for obtaining reflectograms have been discussed. Also, limitations of the \ac{TDR} system based on its parametrization and multiple access schemes have been addressed.

The carried out discussion has been supported by numerical results, which has shown the superiority of the channel estimation procedure over the pulse compression. Furthermore, the influence of the \ac{TDR} system parametrization on sidelobe level, range resolution, and maximum unambiguous range. The attained results show that the \ac{NB}-\ac{PLC} frequency range is suitable for the sensing of \ac{LV} and \ac{MV} power distribution networks, with the appropriate frequency bandwidth for monitoring a power distribution network section being inversely proportional to its length. For the latter analysis, typical European underground \ac{LV} and US overhead \ac{MV} scenarios have been considered and compliance to \ac{FCC}, \ac{ARIB}, and \ac{CENELEC} \ac{NB}-\ac{PLC} regulations has also been addressed.

Finally, a comparison among multiple access schemes has been performed considering a Brazilian overhead \ac{MV} scenario. It has been shown that \ac{TDMA} offers a fair \ac{SINR} level for all \acp{PLM} sensing a power distribution network, while the use of \ac{FDMA} results in higher \ac{SINR} to \acp{PLM} associated with subcarriers in higher frequnecy bins due to the exponentially decreasing additive noise \ac{PSD}. The attained \ac{SINR} levels by the \acp{PLM} in \ac{CDMA} scheme is higher than in the other schemes and fair among the \acp{PLM}, which is due to the averaging process that lowers the additive noise \ac{PSD} in the decoding stage. Additionally, a higher number of reflectograms, which may be distorted if the reflection channel coherence bandwidth constraint is not observed, is obtained over time in the \ac{FDMA} scheme. \ac{TDMA} and \ac{CDMA} schemes, however, obtain less but non-distorted reflectograms due to time multiplexing and spreading processes.

\bibliographystyle{IEEEtran}
\bibliography{referencias}

\begin{thebibliography}{10}
\providecommand{\url}[1]{#1}
\csname url@samestyle\endcsname
\providecommand{\newblock}{\relax}
\providecommand{\bibinfo}[2]{#2}
\providecommand{\BIBentrySTDinterwordspacing}{\spaceskip=0pt\relax}
\providecommand{\BIBentryALTinterwordstretchfactor}{4}
\providecommand{\BIBentryALTinterwordspacing}{\spaceskip=\fontdimen2\font plus
\BIBentryALTinterwordstretchfactor\fontdimen3\font minus
  \fontdimen4\font\relax}
\providecommand{\BIBforeignlanguage}[2]{{%
\expandafter\ifx\csname l@#1\endcsname\relax
\typeout{** WARNING: IEEEtran.bst: No hyphenation pattern has been}%
\typeout{** loaded for the language `#1'. Using the pattern for}%
\typeout{** the default language instead.}%
\else
\language=\csname l@#1\endcsname
\fi
#2}}
\providecommand{\BIBdecl}{\relax}
\BIBdecl

\bibitem{sedighizadeh2010}
M.~Sedighizadeh, A.~Rezazadeh, and N.~I. Elkalashy, ``Approaches in high
  impedance fault detection - a chronological review,'' \emph{Advances in
  Electrical and Computer Engineering}, vol.~10, no.~3, pp. 114--128, Aug 2010.

\bibitem{ghaderi2017}
A.~Ghaderi, H.~L. Ginn, and H.~A. Mohammadpour, ``High impedance fault
  detection: A review,'' \emph{Electric Power Systems Research}, vol. 143, pp.
  376 -- 388, Feb. 2017.

\bibitem{auzanneau2013}
F.~Auzanneau, ``Wire troubleshooting and diagnosis: Review and perspectives,''
  \emph{Progress In Electromagnetics Research B}, vol.~49, pp. 253--279, 2013.

\bibitem{passerini2018_2}
\BIBentryALTinterwordspacing
F.~Passerini and A.~M. Tonello, ``Smart grid monitoring using power line
  modems: Anomaly detection and localization,'' \emph{Cornell University
  Library}, pp. 1--8, July 2018. [Online]. Available:
  \url{https://arxiv.org/abs/1807.05347}
\BIBentrySTDinterwordspacing

\bibitem{ahmed2013}
M.~O. Ahmed and L.~Lampe, ``Power line communications for low-voltage power
  grid tomography,'' \emph{IEEE Transactions on Communications}, vol.~61,
  no.~12, pp. 5163--5175, Dec. 2013.

\bibitem{milioudis2015}
A.~N. Milioudis, G.~T. Andreou, and D.~P. Labridis, ``Detection and location of
  high impedance faults in multiconductor overhead distribution lines using
  power line communication devices,'' \emph{IEEE Transactions on Smart Grid},
  vol.~6, no.~2, pp. 894--902, March 2015.

\bibitem{auzanneau2016}
F.~Auzanneau, ``Transferometry: A new tool for complex wired networks
  diagnosis,'' \emph{Progress In Electromagnetics Research B}, vol.~70, pp.
  87--100, 2016.

\bibitem{furse2006}
C.~Furse, Y.~Chung, C.~Lo, and P.~Pendayala, ``A critical comparison of
  reflectometry methods for location of wiring faults,'' \emph{Smart Structures
  and Systems}, vol.~2, no.~1, pp. 25--46, Jan. 2006.

\bibitem{wang2010}
J.~Wang, P.~E.~C. Stone, Y.~. Shin, and R.~A. Dougal, ``Application of joint
  time-frequency domain reflectometry for electric power cable diagnostics,''
  \emph{IET Signal Processing}, vol.~4, no.~4, pp. 395--405, Aug. 2010.

\bibitem{taylor1996}
V.~Taylor and M.~Faulkner, ``Line monitoring and fault location using spread
  spectrum on power line carrier,'' \emph{IEE Proceedings - Generation,
  Transmission and Distribution}, vol. 143, no.~5, pp. 427--434, Sept. 1996.

\bibitem{bo1999}
Z.~Q. Bo, G.~Weller, and M.~A. Redfern, ``Accurate fault location technique for
  distribution system using fault-generated high-frequency transient voltage
  signals,'' \emph{IEE Proceedings - Generation, Transmission and
  Distribution}, vol. 146, no.~1, pp. 73--79, Jan 1999.

\bibitem{paulis2016}
F.~de~Paulis, C.~Olivieri, A.~Orlandi, and G.~Giannuzzi, ``Detectability of
  degraded joint discontinuities in {HV} power lines through {TDR}-like remote
  monitoring,'' \emph{IEEE Transactions on Instrumentation and Measurement},
  vol.~65, no.~12, pp. 2725--2733, Dec 2016.

\bibitem{hassen2015}
W.~B. Hassen, F.~Auzanneau, L.~Incarbone, F.~P\'er\`es, and A.~P. Tchangani,
  ``Distributed sensor fusion for wire fault location using sensor clustering
  strategy,'' \emph{International Journal of Distributed Sensor Networks},
  vol.~11, no.~4, pp. 1--17, April 2015.

\bibitem{naik2006}
S.~Naik, C.~M. Furse, and B.~Farhang-Boroujeny, ``Multicarrier reflectometry,''
  \emph{IEEE Sensors Journal}, vol.~6, no.~3, pp. 812--818, June 2006.

\bibitem{amini2009}
P.~Amini, C.~Furse, and B.~Farhang-Boroujeny, ``Filterbank multicarrier
  reflectometry for cognitive live wire testing,'' \emph{IEEE Sensors Journal},
  vol.~9, no.~12, pp. 1831--1837, Dec. 2009.

\bibitem{lelong2009}
A.~Lelong and M.~O. Carrion, ``On line wire diagnosis using multicarrier time
  domain reflectometry for fault location,'' in \emph{Proc. IEEE SENSORS}, Oct.
  2009, pp. 751--754.

\bibitem{oliveira2014}
T.~R. Oliveira, C.~A.~G. Marques, W.~A. Finamore, S.~L. Netto, and M.~V.
  Ribeiro, ``A methodology for estimating frequency responses of electric power
  grids,'' \emph{Journal of Control, Automation and Electrical Systems},
  vol.~25, no.~6, pp. 720--731, Dec. 2014.

\bibitem{chang2015}
S.~J. Chang, C.~K. Lee, C.~Lee, Y.~J. Han, M.~K. Jung, J.~B. Park, and Y.~Shin,
  ``Condition monitoring of instrumentation cable splices using kalman
  filtering,'' \emph{IEEE Transactions on Instrumentation and Measurement},
  vol.~64, no.~12, pp. 3490--3499, Dec. 2015.

\bibitem{lelong2010}
A.~Lelong, L.~Sommervogel, N.~Ravot, and M.~O. Carrion, ``Distributed
  reflectometry method for wire fault location using selective average,''
  \emph{IEEE Sensors Journal}, vol.~10, no.~2, pp. 300--310, Feb. 2010.

\bibitem{paul2007}
C.~R. Paul, \emph{Analysis of Multiconductor Transmission Lines, 2nd
  Edition}.\hskip 1em plus 0.5em minus 0.4em\relax John Wiley \& Sons Inc.,
  2007.

\bibitem{giroto2018}
L.~G. de~Oliveira, G.~R. Colen, A.~J.~H. Vinck, and M.~V. Ribeiro, ``Resource
  allocation in {HS-OFDM}-based {PLC} systems: A tutorial,'' \emph{Journal of
  Communication and Information Systems}, vol.~33, no.~1, Oct. 2018.

\bibitem{mitra2010}
S.~K. Mitra, \emph{Digital Signal Processing: A Computer-Based Approach, 4th
  Edition}.\hskip 1em plus 0.5em minus 0.4em\relax McGraw-Hill, 2010.

\bibitem{passerini2017}
F.~Passerini and A.~M. Tonello, ``Power line fault detection and localization
  using high frequency impedance measurement,'' in \emph{Proc. IEEE
  International Symposium on Power Line Communications and its Applications
  (ISPLC)}, April 2017, pp. 1--5.

\bibitem{hassen2018}
W.~B. Hassen, M.~Kafal, and E.~Cabanillas, ``Time reversal applied to
  multi-carrier reflectometry for on-line diagnosis in complex wiring
  systems,'' in \emph{Proc. IEEE AUTOTESTCON}, Sept. 2018, pp. 1--7.

\bibitem{sturm2009}
C.~Sturm, E.~Pancera, T.~Zwick, and W.~Wiesbeck, ``A novel approach to {OFDM}
  radar processing,'' in \emph{Proc. IEEE Radar Conference}, May 2009, pp.
  1--4.

\bibitem{parte1}
\BIBentryALTinterwordspacing
L.~G. de~Oliveira, M.~de~L.~Filomeno, L.~F. Colla, H.~V. Poor, and M.~V.
  Ribeiro, ``On the suitability of {PLC} pulses for power line fault sensing
  via time-domain reflectometry,'' \emph{Cornell University Library}, pp.
  1--13, Jan. 2019. [Online]. Available: \url{https://arxiv.org/abs/1901.07923}
\BIBentrySTDinterwordspacing

\bibitem{fitch1988}
J.~P. Fitch, \emph{Synthetic Aperture Radar}, C.~Burrus, Ed.\hskip 1em plus
  0.5em minus 0.4em\relax Springer-Verlag, 1988.

\bibitem{lellouch2016}
G.~Lellouch, A.~K. Mishra, and M.~Inggs, ``Design of {OFDM} radar pulses using
  genetic algorithm based techniques,'' \emph{IEEE Transactions on Aerospace
  and Electronic Systems}, vol.~52, no.~4, pp. 1953--1966, Aug. 2016.

\bibitem{richardsvol1}
M.~A. Richards, J.~A. Scheer, and W.~A. Holm, \emph{Principles of Modern Radar
  - Basic Principles}.\hskip 1em plus 0.5em minus 0.4em\relax Scitech
  Publishing Inc., 2010, vol.~1.

\bibitem{temes1962}
C.~L. Temes, ``Sidelobe suppression in a range-channel pulse-compression
  radar,'' \emph{IRE Transactions on Military Electronics}, vol. MIL-6, no.~2,
  pp. 162--169, April 1962.

\bibitem{sit2017}
Y.~L. Sit, ``{MIMO} {OFDM} radar-communication system with mutual interference
  cancellation,'' Ph.D. dissertation, Karlsruhe Institute of Technology,
  Germany, 2017.

\bibitem{colenSCRA}
G.~R. Colen, L.~G. de~Oliveira, A.~J.~H. Vinck, and M.~V. Ribeiro, ``A spectral
  compressive resource allocation technique for {PLC} systems,'' \emph{IEEE
  Transactions on Communications}, vol.~65, no.~2, pp. 816--826, Feb. 2017.

\bibitem{nuss2018}
B.~Nuss, J.~Mayer, and T.~Zwick, ``Limitations of {MIMO} and multi-user access
  for {OFDM} radar in automotive applications,'' in \emph{Proc. IEEE MTT-S
  International Conference on Microwaves for Intelligent Mobility}, April 2018,
  pp. 1--4.

\bibitem{hassen_tese2014}
W.~B. Hassen, ``\'{E}tude de strat\'egies de diagnostic embarqu\'e des
  r\'eseaux filaires complexes,'' Ph.D. dissertation, University of Toulouse,
  France, 2014, in French.

\bibitem{filomeno2018}
M.~de~L.~Filomeno, G.~R. Colen, L.~G. de~Oliveira, and M.~V. Ribeiro,
  ``Two-stage single-relay channel model for in-home broadband {PLC} systems,''
  \emph{IEEE Systems Journal (Early Access)}, pp. 1--11, July 2018.

\bibitem{sturm2013}
C.~Sturm, Y.~L. Sit, M.~Braun, and T.~Zwick, ``Spectrally interleaved
  multi-carrier signals for radar network applications and multi-input
  multi-output radar,'' \emph{IET Radar, Sonar Navigation}, vol.~7, no.~3, pp.
  261--269, March 2013.

\bibitem{1901_2}
``{IEEE} standard for low-frequency (less than 500 {kHz}) narrowband power line
  communications for smart grid applications,'' \emph{{IEEE} Std 1901.2-2013},
  pp. 1--269, Dec. 2013.

\bibitem{lampe_vinck2011}
L.~Lampe and A.~J.~H. Vinck, ``On cooperative coding for narrow band {PLC}
  networks,'' \emph{AEU - International Journal of Electronics and
  Communications}, vol.~65, no.~8, pp. 681--687, Aug. 2011.

\bibitem{PLCbook2016}
L.~Lampe, A.~M. Tonello, and T.~G. Swart, \emph{Power Line Communications:
  Principles, Standards and Applications from Multimedia to Smart Grid, 2nd
  Edition}.\hskip 1em plus 0.5em minus 0.4em\relax John Wiley \& Sons Inc.,
  2016.

\bibitem{franek2017}
L.~Franek and P.~Fiedler, ``A multiconductor model of power line communication
  in medium-voltage lines,'' \emph{Energies}, vol.~10, no.~6, pp. 1--16, 2017.

\bibitem{tao2007}
Z.~Tao, Y.~Xiaoxian, Z.~Baohui, N.~H. Xu, F.~Xiaoqun, and L.~Changxin,
  ``Statistical analysis and modeling of noise on 10-kv medium-voltage power
  lines,'' \emph{IEEE Transactions on Power Delivery}, vol.~22, no.~3, pp.
  1433--1439, July 2007.

\bibitem{girotto2017}
M.~Girotto and A.~M. Tonello, ``{EMC} regulations and spectral constraints for
  multicarrier modulation in {PLC},'' \emph{IEEE Access}, vol.~5, pp.
  4954--4966, Mar. 2017.

\end{thebibliography}

\end{document}